\documentstyle[psfig,aps,pra,multicol]{revtex}
\begin{document}

\title{The influence of density of modes on dark lines
in spontaneous emission}

\author{E. Paspalakis, D.G. Angelakis and  P.L. Knight}

\address{Optics Section, Blackett Laboratory, Imperial College, 
London SW7 2BZ, England}

\date{\today}

\maketitle
\begin{abstract}
We study two distinct multi-level 
atomic models in which one transition is coupled to a Markovian
reservoir, while another linked transition is coupled to a non-Markovian
reservoir. We show that by choosing appropriately the density of modes
of the non-Markovian reservoir the spontaneous emission to the
Markovian reservoir is greatly altered. The existence
of `dark lines' in the spontaneous emission spectrum
in the Markovian reservoir due to the coupling to specific
density of modes of the non-Markovian reservoir is also
predicted.
\end{abstract}



\section{Introduction}
It is now well understood that spontaneous emission
of a quantum system depends crucially on the nature of the reservoir 
with which the
system interacts. Spontaneous emission can be modified
in a ``tailored'' manner
by changing the density of modes of the reservoir \cite{CQED}. 
An  effective method to achieve this is to place atoms in waveguides \cite{Kleppner81a,Lewenstein88a,Frerichs95a}, microcavities \cite{Lewenstein87a,Lai88a,Keitel95a,Rippin96a,Garraway96a,Buzek99a} or
 photonic band gap materials \cite{Nabiev93a,John94a,Kurizki94a,Quang97a,Zhu97a,Bay97a,Bay98a,Vats98a},
where the density of modes differs substantially from that of
the free space vacuum.
The latter has also attracted attention 
for its potential for modifying the absorption and dispersion properties 
of a system \cite{Paspalakis99a}.

In the above studies the typical scheme involves the interaction of a two-level
atom with a reservoir with modified density of modes. 
Spontaneous emission of this two-level system
differs considerably from the free space result and
the usual Weisskopf-Wigner exponential decay \cite{Books1} is violated.  
In specific cases complete inhibition of spontaneous decay
has even been predicted \cite{John94a,Kurizki94a}. In addition
to studies of two-level atoms, 
there are also investigations of multi-level atoms, where 
the spontaneous emission from an atomic transition
in the modified reservoir influences the spontaneous emission of another
atomic transition that interacts with the normal free space vacuum \cite{John94a,Bay97a,Bay98a}. It is schemes such as these 
that concerns us in this article. 
Specifically, we study two distinct multi-level atomic schemes
where one transition is considered to decay
spontaneously in a modified reservoir
and the other decays to a normal free space vacuum. The main interest here
is the existence of `dark lines' (complete
quenching of spontaneous emission for specific vacuum modes)
in the spontaneous emission spectrum of the {\it free space}
transition due to the coupling of the other transition
with the modified reservoir. We note that dark lines
in spontaneous emission have been predicted in several
laser driven schemes \cite{Zhu95a,Toor95a,Agarwal98a,Paspalakis98a}
and have been experimentally observed \cite{Xia96a}.
In this article we employ specific models for the modified reservoir
density of modes, such as, for example, 
that which results from an isotropic 
photonic band gap model with (or without) defects. We show that
dark lines appear in the spontaneous emission
spectrum as a consequence of the structure of the modified
reservoir.

This article is organized as follows: in the next section we consider
a three-level, $\Lambda$-type atomic system, 
with one transition spontaneously decaying 
in the normal free space vacuum and the other transition 
decaying in a modified reservoir.
We study the spontaneous emission spectrum of the free space transition
for four different density of modes of the modified reservoir and show
that dark lines can occur in the spectrum due to the structure of
the modified reservoir. We note that John and Quang \cite{John94a} 
have studied the 
same atomic system as us, using specifically the appropriate
density of modes obtained near the edge of an isotropic 
photonic band gap [see Eq. (\ref{isodom})]. However, in their study they 
focussed on the phenomenon of dynamical splitting in the spectrum and 
not on the existence dark lines, which is the main phenomenon 
discussed in our article. Zeros and splittings in spectra should not 
be confused, although they relate to each other in the limit of 
strong coupling. In section III we consider a laser-driven
extension of the previous system and show that dark lines can occur in this
system, too. In this case the dark lines originate from either laser-induced
or modified vacuum-induced mechanisms. Finally, we summarize 
our findings in section IV.

\section{First case: $\Lambda$-type scheme}

\begin{figure}
\centerline{\psfig{figure=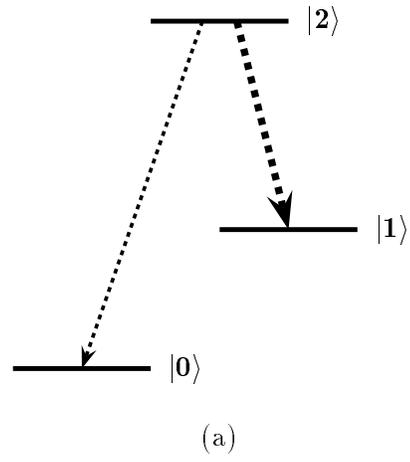,height=6.0cm}}

\vspace*{1.5cm}

\centerline{\psfig{figure=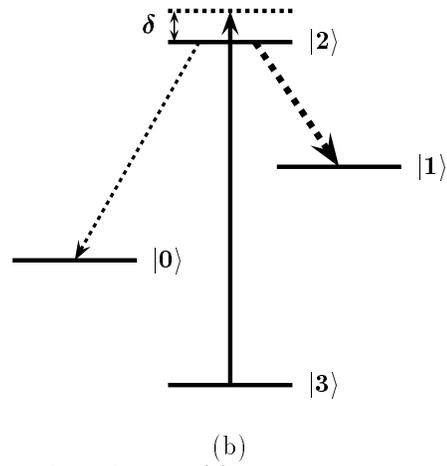,height=6.0cm}}
\caption
{\narrowtext The two systems under consideration. 
In (a) the 
thick dashed line
denotes the coupling to the modified reservoir
and the thin dashed line denotes the coupling to 
the Markovian reservoir. The same hold in (b), and in addition
the solid line denotes the coupling by the laser field.}
\label{fig1}
\end{figure}

We begin with the study of the $\Lambda$-type scheme, shown in Fig.\ \ref{fig1}(a).
This system is similar to that used by Lewenstein {\it et al} \cite{Lewenstein88a} and by John and Quang
\cite{John94a}.
The atom is assumed to be initially in state $|2\rangle$.
The transition $|2\rangle \leftrightarrow |1\rangle$ is taken 
to be near resonant with a modified reservoir (this will
be later referred to as the non-Markovian reservoir), while
the transition $|2\rangle \leftrightarrow |0\rangle$ is assumed to be 
occurring in free space (this will
be later referred to as the Markovian reservoir). The spectrum of this latter transition is of central interest in this article.
The Hamiltonian which describes the 
dynamics of this system, in the interaction picture and the rotating wave approximation (RWA), is given by
(we use units such that $\hbar = 1$),
\begin{eqnarray} 
H &=& \sum_{{\bf \lambda}}g_{{\bf \lambda}}e^{-i(\omega_{\bf \lambda}- \omega_{20})t} |2\rangle \langle 0| a_{{\bf \lambda}} +
\sum_{{\bf \kappa}}g_{{\bf \kappa}}e^{-i(\omega_{\bf \kappa}- \omega_{21})t} |2\rangle \langle 1| a_{{\bf \kappa}} 
+ \mbox{H.c.}  \, . \label{Ham}
\end{eqnarray}
Here, $g_{{\bf \kappa}}$  
denotes the coupling of the atom with the 
modified vacuum modes $({\bf \kappa})$ and $g_{{\bf \lambda}}$  
denotes the coupling of the atom with the 
free space vacuum modes $({\bf \lambda})$. Both coupling
strengths are taken to be real. The energy separations of the states
are denoted by $\omega_{ij} = \omega_{i} - \omega_{j}$ 
and $\omega_{\bf \kappa}$ $(\omega_{\bf \lambda})$ is the energy of
the ${\bf \kappa}$ $({\bf \lambda})$-th reservoir mode.

The description of the system is given using a probability amplitude approach.
We proceed by expanding the wave function of the system, 
at a specific time $t$, in terms of the
`bare' state vectors such that
\begin{eqnarray}
 |\psi(t)\rangle = b_{2}(t)|2,\{0\}\rangle 
 + \sum_{{\bf \lambda}}b_{{\bf \lambda}}(t)|0,\{{\bf \lambda}\}\rangle
+ \sum_{{\bf \kappa}}b_{{\bf \kappa}}(t)|1,\{{\bf \kappa}\}\rangle
 \, . \label{wav}
\end{eqnarray}
Substituting Eqs.\ (\ref{Ham}) and (\ref{wav}) into the time-dependent Schr\"{o}dinger equation we obtain 
\begin{eqnarray}
i\dot{b}_{2}(t) &=& \sum_{\bf \lambda} g_{{\bf \lambda}} b_{\bf \lambda} (t) e^{-i(\omega_{{\bf \lambda}} - \omega_{20}) t} + 
\sum_{\bf \kappa} g_{{\bf \kappa}} b_{\bf \kappa} (t) e^{-i(\omega_{{\bf \kappa}} - \omega_{21}) t} \label{21} \, , \\
i\dot{b}_{\bf \lambda}(t) &=& g_{{\bf \lambda}} b_{2}(t) e^{i(\omega_{{\bf \lambda}} - \omega_{20}) t} \label{2l} \, , \\
i\dot{b}_{\bf \kappa}(t) &=& g_{{\bf \kappa}} b_{2}(t) e^{i(\omega_{{\bf \kappa}} - \omega_{21}) t} \label{2k} \, .
\end{eqnarray}
We proceed by performing a formal time integration of Eqs.\ (\ref{2l}) and (\ref{2k}) and substitute the result into Eq.\ (\ref{21}) to obtain 
the integro-differential equation
\begin{eqnarray}
\dot{b}_{2}(t) &=&  
-\int^{t}_{0}dt^{\prime}b_{2} (t^{\prime})\sum_{\bf \lambda} g^{2}_{{\bf \lambda}}  e^{-i(\omega_{{\bf \lambda}} - \omega_{20}) (t-t^{\prime})}
- \int^{t}_{0}dt^{\prime}b_{2} (t^{\prime})\sum_{\bf \kappa} g^{2}_{{\bf \kappa}}  e^{-i(\omega_{{\bf \kappa}} - \omega_{20}) (t-t^{\prime})}  \label{21a} \, .
\end{eqnarray}
Because the reservoir with modes $\lambda$ is assumed 
to be Markovian, we can apply the usual 
Weisskopf-Wigner result \cite{Books1} and obtain
\begin{equation}
\sum_{\bf \lambda} g^{2}_{{\bf \lambda}}  e^{-i(\omega_{{\bf \lambda}} - \omega_{20}) (t-t^{\prime})} = \frac{\gamma}{2}\delta(t - t^{\prime}) \, .
\label{weisskopf}
\end{equation}
Note that the principal value term associated with the Lamb shift 
which should accompany the decay rate has
been omitted in Eq.\ (\ref{weisskopf}). This does
not affect our the results, as we can assume that the Lamb shift 
is incorporated into the definition of our state energies. For the second summation in Eq.\ (\ref{21a}),
the one associated with the modified reservoir modes, the above result
is not applicable as the density of modes of this reservoir is assumed
to vary much quicker than that of free space. To tackle this problem, we 
define the following kernel
\begin{equation}
K(t-t^{\prime}) = \sum_{{\bf \kappa}}  g^{2}_{{\bf \kappa}} e^{-i(\omega_{\bf \kappa}- \omega_{21}) (t-t^{\prime})} \approx 
g^{3/2}\int d\omega \rho(\omega) e^{-i(\omega - \omega_{21}) (t-t^{\prime})} \, . \label{kerneltot}
\end{equation}
which is calculated using the appropriate
density of modes $\rho(\omega)$ of the 
modified reservoir.
In Eq.\ (\ref{kerneltot}), $g$ denotes the coupling constant 
of the atom to the non-Markovian reservoir. Using Eqs.\ (\ref{weisskopf})
and (\ref{kerneltot}) into Eq.\ (\ref{21a}) we obtain
\begin{eqnarray}
\dot{b}_{2}(t) &=&  
-\frac{\gamma}{2} b_{2} (t)
- \int^{t}_{0}dt^{\prime}b_{2} (t^{\prime}) K(t-t^{\prime})  \label{21a1} \, .
\end{eqnarray}

The long time spontaneous emission spectrum in the Markovian reservoir
is given by 
$S(\delta_{\bf \lambda}) \propto |b_{\bf \lambda}(t \rightarrow \infty)|^2$,
with $\delta_{\bf \lambda} = \omega_{\bf \lambda} - \omega_{20}$ 
\cite{Zhu95a,Paspalakis98a}.
We calculate $b_{\bf \lambda}(t \rightarrow \infty)$ with the use of
the Laplace transform \cite{Barnettbook} of the equations of
motion. Using Eq.\ (\ref{2l}) and the final
value theorem \cite{Barnettbook} we obtain the spontaneous emission
spectrum as
$S(\delta_{{\bf \lambda}}) \propto {\gamma} |\lim_{s \rightarrow -i \delta_{{\bf \lambda}}} B_{2}(s)|^{2}$, where $B_{2}(s)$ is the Laplace transform of the atomic amplitudes $b_{2}(t)$ and $s$ is the Laplace variable. This in turn, 
with the help of Eq.\ (\ref{21a1}), reduces to
\begin{equation}
S(\delta_{{\bf \lambda}}) \propto \frac{\gamma}{\left|-i \delta_{\bf \lambda} + \gamma/2 + \tilde{K}(s \rightarrow -i \delta_{\bf \lambda})\right|^2} \, ,
\end{equation}
where $\tilde{K}(s)$ is the Laplace transform of $K(t)$, which yields
\begin{equation}
\tilde{K}(s) =  g^{3/2} \int d\omega \frac{\rho(\omega)}{s + i (\omega - \omega_{21})} \, . \label{kernellap}
\end{equation}
Therefore, in order to calculate the spontaneous emission spectrum in the Markovian reservoir
we need to calculate $\tilde{K}(s)$. This will be done for different models of the density of
modes $\rho(\omega)$ of the non-Markovian reservoir. 

\begin{figure}
\centerline{\psfig{figure=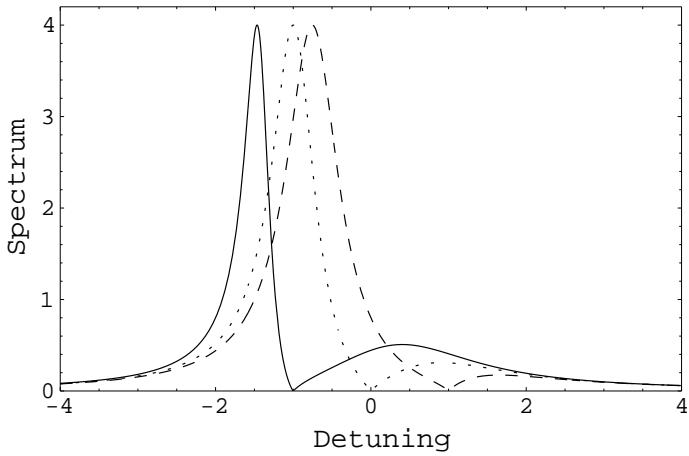,height=6.0cm}}
\caption
{\narrowtext The spontaneous emission spectrum $S(\delta_{\bf \lambda})$ (in arbitrary
units) given by Eq.\ (\ref{isospec})
for parameters 
$g = 1$, and $\delta_{g} = 0$ (dotted curve); 
$\delta_{g} = 1$ (dashed curve); $\delta_{g} = -1$ (full curve). 
All parameters are in units of $\gamma$. }
\label{fig2}
\end{figure}

We begin by considering the non-Markovian
reservoir to be that obtained near the edge of an isotropic photonic band gap model \cite{John94a,Kurizki94a}. Then, 
\begin{equation}
\rho(\omega) = \frac{1}{\pi} \frac{1}{\sqrt{\omega - \omega_{g}}}\Theta(\omega - \omega_{g}) \, , \label{isodom}
\end{equation}
where $\omega_{g}$ is the gap frequency and $\Theta$ is the Heaviside step function. 
In this case, Eq.\ (\ref{kernellap}) leads to 
\begin{equation}
\tilde{K}(s) = \frac{g^{3/2}}{\sqrt{is - \delta_{g}}} \, , \label{isoker}
\end{equation}
with $\delta_{g} = \omega_{g} - \omega_{21}$. The spontaneous emission
spectrum then reads
\begin{equation}
S(\delta_{\bf \lambda}) \propto  \gamma \left|\frac{\sqrt{\delta_{\bf \lambda} - \delta_{g}}}{(-i\delta_{\bf \lambda}+\gamma/2)\sqrt{\delta_{\bf \lambda} - \delta_{g}} + g^{3/2}} \right|^{2}\, . \label{isospec}
\end{equation}
Obviously, the spectrum exhibits a zero (i.e. predicts the
existence of a dark line), if $\delta_{\bf \lambda} = \delta_{g}$. This is purely an effect
of the above density of states, and the non-Markovian character
of the reservoir. In the case of a Markovian reservoir the
spontaneous emission spectrum would obtain the well-known
Lorentzian profile and no dark line would appear in the spectrum. 
The behaviour of the spectrum is shown in Fig.\ \ref{fig2} 
for different values
of the detuning from the threshold.
The spectrum has two well-separated peaks and the dark line appears
at the predicted value.
We note that a spectrum of this form has also been derived
by John and Quang \cite{John94a}; however, in their study they 
did not mention the existence of the dark line 
(i.e. the zero in the spectrum and its spectral position),
which is of central importance in this article.

\begin{figure}
\centerline{\psfig{figure=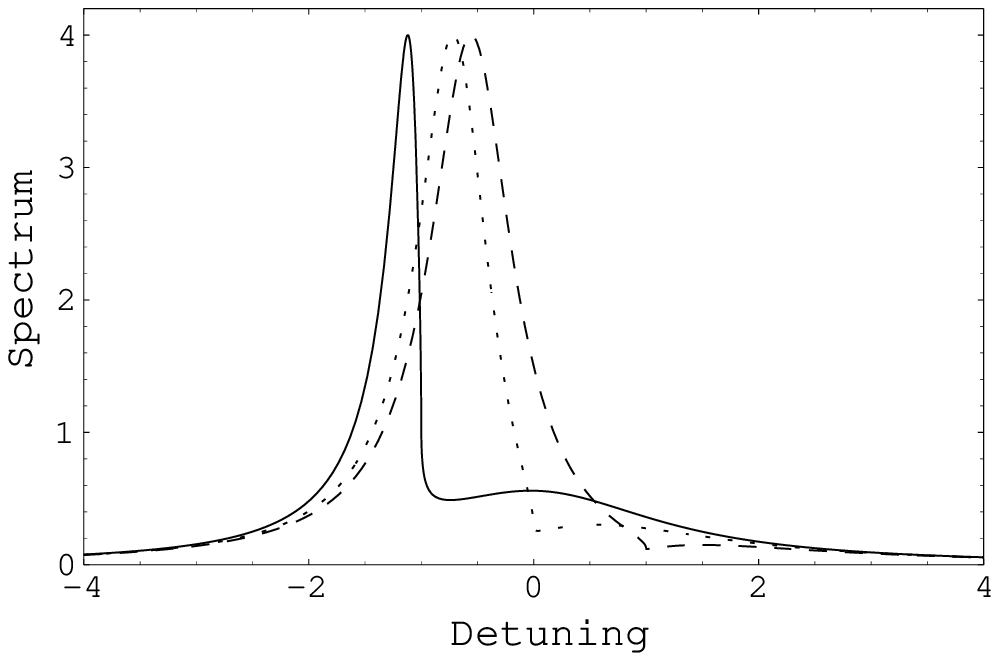,height=6.0cm}}
\caption
{\narrowtext The spontaneous emission spectrum $S(\delta_{\bf \lambda})$ given by Eq.\ (\ref{smospec})
for parameters 
$g = 1$, $\epsilon = 0.3$, and $\delta_{g} = 0$ (dotted curve); 
$\delta_{g} = 1$ (dashed curve); $\delta_{g} = -1$ (full curve). }
\label{fig3}
\end{figure}

The density of modes of Eq.\ (\ref{isodom}) 
is a special case of a more general family
of density of modes those given by
\begin{equation}
\rho(\omega) = \frac{1}{\pi}\frac{\sqrt{\omega - \omega _{g}}}{\epsilon + {\omega - \omega_{g}}}\Theta(\omega - \omega_{g}) \, , \label{smodom}
\end{equation}
where $\epsilon$ is usually referred to as the smoothing parameter. Such a density of modes
has been used in studies of atoms in waveguides \cite{Kleppner81a,Lewenstein88a} microcavities \cite{Lewenstein87a,Rippin96a} and photonic band gap materials \cite{Kurizki94a}. Eq.\  (\ref{isodom})
is recovered by taking the limit $\epsilon \rightarrow 0$ in Eq.\  (\ref{smodom}).  For the above
density of modes, Eq.\ (\ref{smodom}) we obtain
\begin{eqnarray}
\tilde{K}(s)& =& \frac{g^{3/2}}{i\sqrt{\epsilon} + \sqrt{is - \delta_{g}}} \, , \\
S(\delta_{\bf \lambda}) &\propto&  \gamma \left|\frac{i\sqrt{\epsilon}+\sqrt{\delta_{\bf \lambda} - \delta_{g}}}{(-i\delta_{\bf \lambda}+\gamma/2)(i\sqrt{\epsilon}+\sqrt{\delta_{\bf \lambda} - \delta_{g}}) + g^{3/2}} \right|^{2}\, . \label{smospec}
\end{eqnarray}
So, in this ``smoothed'' case, no dark line appears in the spectrum. This
is shown in Fig.\ \ref{fig3}, where the zero disappears from the spectrum.
The only case for which spontaneous 
emission spectrum will give zero is that of $\epsilon = 0$ and $\delta_{\bf \lambda} = \delta_{g}$ which, of course, reduces to the previous
result of Eq.\ (\ref{isospec}).

The above two density of modes describe ``pure'' materials
(i.e. materials with no defects). In reality, defects exist
in waveguides, microcavities or photonic band gap materials which exhibit gaps in their density of modes. These defects lead to narrow-linewidth, well-localized modes in the gaps of the
above structures. Their density of modes can be described by either a delta function
\cite{Nabiev93a} or a narrow Lorentzian \cite{Kurizki94a}. We will now combine the density of modes given by Eq.\ (\ref{isodom})
and that given by a defect structure. In the case that we choose a delta function density
of modes for the defect structure we obtain
\begin{eqnarray}
\tilde{K}(s)& =& \frac{g^{3/2}}{ \sqrt{is - \delta_{g}}} + \frac{g^{2}_{1}}{s + i \delta_{c}}    \, , \\
S(\delta_{\bf \lambda}) &\propto&  \gamma \left|\frac{\sqrt{\delta_{\bf \lambda} - \delta_{g}}}{(-i\delta_{\bf \lambda}+\gamma/2)\sqrt{\delta_{\bf \lambda} - \delta_{g}} + g^{3/2}+  ig^{2}_{1}\frac{\sqrt{\delta_{\bf \lambda} - \delta_{g}}}{\delta_{\bf \lambda}- \delta_{c}}} \right|^{2}\, , \label{deltaspec}
\end{eqnarray}
where $\delta_{c} = \omega_{c} - \omega_{21}$ is the defect mode-atom detuning (defect mode at frequency $\omega_{c}$) and $g_{1}$ is the defect mode-atom coupling constant. In this case the
spectrum exhibits two dark lines one at $\delta_{\bf \lambda} = \delta_{g}$ and another
at $\delta_{\bf \lambda} = \delta_{c}$. In Fig.\ \ref{fig4} we present
the spontaneous emission spectrum described by Eq.\ (\ref{deltaspec})
for the same parameters
that used in Fig.\ \ref{fig2}. The spectrum obtains now a very
pronounced third peak and two dark lines at the predicted values. 
For a Lorentzian density of modes of the defect structure we get
 \begin{eqnarray}
\tilde{K}(s)& =& \frac{g^{3/2}}{ \sqrt{is - \delta_{g}}} + \frac{g^{2}_{1}}{s + i \delta_{c} + \gamma_{c}/2}    \, , \\
S(\delta_{\bf \lambda}) &\propto& \gamma \left|\frac{\sqrt{\delta_{\bf \lambda} - \delta_{g}}}{(-i\delta_{\bf \lambda}+\gamma/2)\sqrt{\delta_{\bf \lambda} - \delta_{g}} + g^{3/2}+  g^{2}_{1}\frac{\sqrt{\delta_{\bf \lambda} - \delta_{g}}}{i(\delta_{c}- \delta_{\bf \lambda}) + \gamma_{c}/2}} \right|^{2}\, ,\label{lorspec}
\end{eqnarray}
with $\gamma_{c}$ being the width of the Lorentzian describing the defect mode. As in the case
of the density of modes given by Eq.\ (\ref{isodom}), in this case too,
there is a single
dark line in the spectrum at $\delta_{\bf \lambda} = \delta_{g}$. 
However, the behaviour of the
spectrum is quite different in this case compared to that shown in Fig.\ \ref{fig2}, as can be seen in Fig.\ \ref{fig5}. 

\section{Second case: Laser-driven scheme}

In this section we turn to the study of the laser-driven system shown
in Fig.\ \ref{fig1}(b). This system is composed of a 
ground state $|3 \rangle$ which is coupled by a laser field to the excited 
state $|2 \rangle$. State $|2 \rangle$
can spontaneously couple to either state $|0 \rangle$ via interaction with 
a Markovian reservoir, or to state $|1 \rangle$ via interaction with 
a non-Markovian reservoir, as in the previous section. 
The spontaneous emission spectrum 
in the Markovian reservoir is our main concern in this section, too.
The Hamiltonian of this system, written in the interaction picture
and under the RWA reads
\begin{eqnarray} 
H &=& \Omega |3\rangle \langle 2| e^{i \delta t} + \sum_{{\bf \lambda}}g_{{\bf \lambda}}e^{-i(\omega_{\bf \lambda}- \omega_{20})t} |2\rangle \langle 0| a_{{\bf \lambda}} +
\sum_{{\bf \kappa}}g_{{\bf \kappa}}e^{-i(\omega_{\bf \kappa}- \omega_{21})t} |2\rangle \langle 1| a_{{\bf \kappa}} 
+ \mbox{H.c.}  \, . \label{2Ham}
\end{eqnarray}
Here, $\Omega$ is the Rabi frequency (assumed real) 
and $\delta = \omega - \omega_{23}$
the laser detuning, with $\omega$ being the 
laser field angular frequency. 
For the description of this system too we use the probability
amplitude approach
and expand the wave function of the system as
\begin{eqnarray}
 |\psi(t)\rangle = b_{3}(t) e^{i \delta t}|3,\{0\}\rangle + b_{2}(t)|2,\{0\}\rangle 
 + \sum_{{\bf \lambda}}b_{{\bf \lambda}}(t)|0,\{{\bf \lambda}\}\rangle
+ \sum_{{\bf \kappa}}b_{{\bf \kappa}}(t)|1,\{{\bf \kappa}\}\rangle
 \, . \label{2wav}
\end{eqnarray}
We substitute Eqs.\ (\ref{2Ham}) and (\ref{2wav}) into the time-dependent Schr\"{o}dinger equation and apply the elimination of the
vacuum modes outlined in the previous section to obtain 
\begin{eqnarray}
i\dot{b}_{3}(t) &=& \delta b_{3}(t) + \Omega b_{2}(t) \label{223} \, , \\
i\dot{b}_{2}(t) &=& \Omega b_{3}(t) -i\frac{\gamma}{2} b_{2} (t)
- i\int^{t}_{0}dt^{\prime}b_{2} (t^{\prime}) K(t-t^{\prime}) \label{221} \, , \\
i\dot{b}_{\bf \lambda}(t) &=& g_{{\bf \lambda}} b_{2}(t) e^{i(\omega_{{\bf \lambda}} - \omega_{20}) t} \label{22l} \, , \\
i\dot{b}_{\bf \kappa}(t) &=& g_{{\bf \kappa}} b_{2}(t) e^{i(\omega_{{\bf \kappa}} - \omega_{21}) t} \label{22k} \, .
\end{eqnarray}

The long time spontaneous emission spectrum in the Markovian reservoir
is given by 
$S(\delta_{\bf \lambda}) \propto |b_{\bf \lambda}(t \rightarrow \infty)|^2$,
and is calculated in a closed form with the use of the Laplace 
transform and the final value theorem \cite{Barnettbook} as
\begin{equation}
S(\delta_{{\bf \lambda}}) \propto \gamma \left|\frac{(\delta_{\bf \lambda} - \delta)b_{2}(0) + \Omega b_{3}(0)}{(\delta_{\bf \lambda} - \delta)[\delta_{\bf \lambda} + i\gamma/2 + i\tilde{K}(s \rightarrow -i \delta_{\bf \lambda})] - \Omega^{2}}\right|^{2} \, .
\end{equation}
The spectrum depends, in this case too, from the Laplace transform
of the kernel, which in turn depends on the density of modes of
the modified reservoir [see Eq.\ (\ref{kernellap})]. We use the four
models of the density of modes described in the previous section to calculate
the above spectrum.

\begin{figure}
\centerline{\psfig{figure=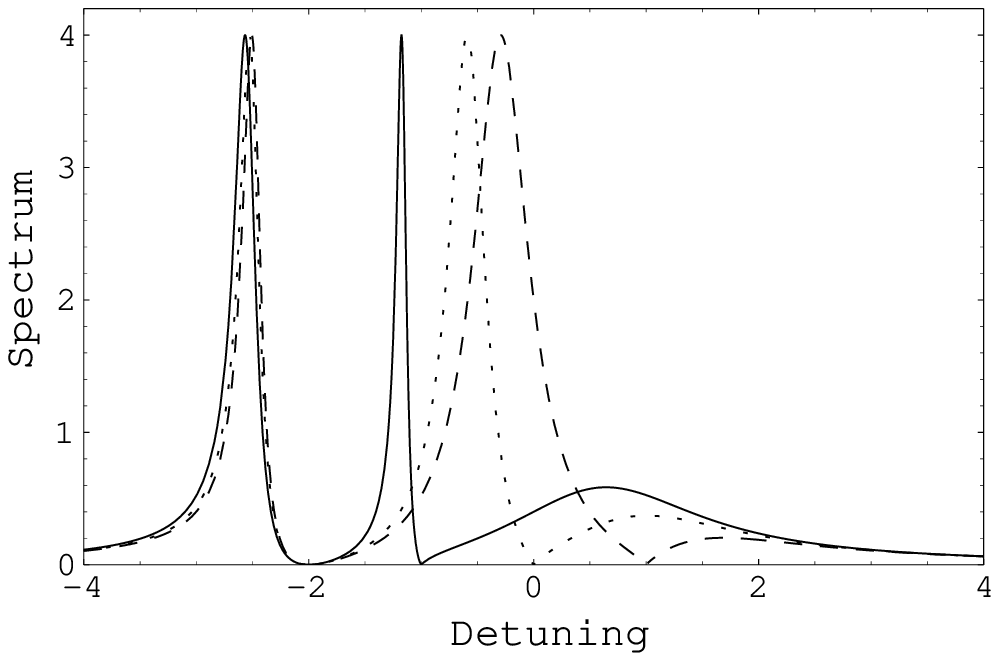,height=6.0cm}}
\caption
{\narrowtext The spontaneous emission spectrum $S(\delta_{\bf \lambda})$ 
given by Eq.\ (\ref{deltaspec})
for parameters 
$g = 1$, $g_{1} =1$, $\delta_{c} = -2$, and $\delta_{g} = 0$ (dotted curve); 
$\delta_{g} = 1$ (dashed curve); $\delta_{g} = -1$ (full curve). }
\label{fig4}
\end{figure}

\begin{figure}
\centerline{\psfig{figure=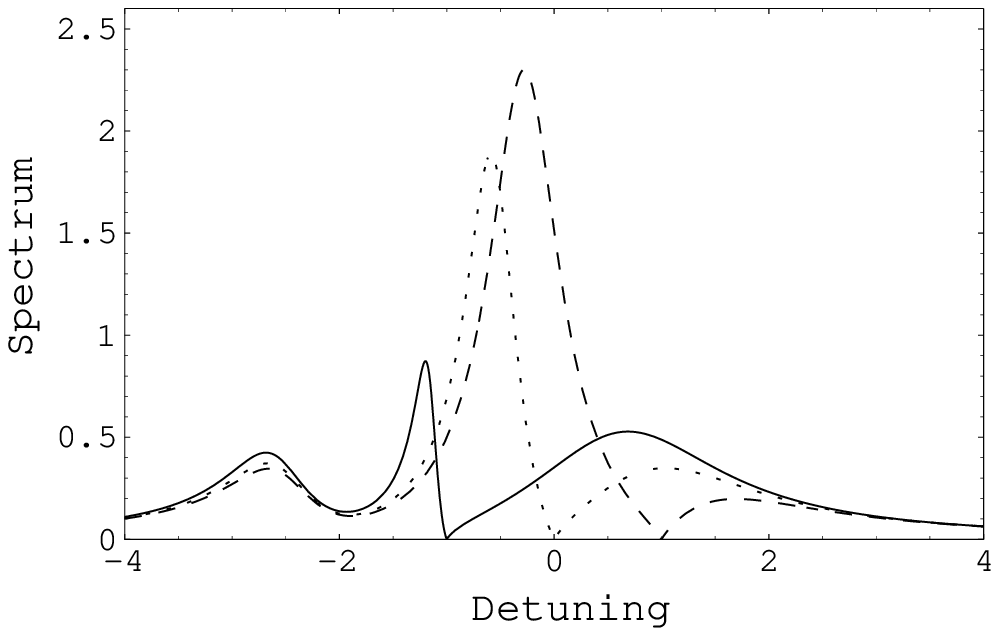,height=6.0cm}}
\caption
{\narrowtext The spontaneous emission spectrum $S(\delta_{\bf \lambda})$ 
given by Eq.\ (\ref{lorspec})
for parameters 
$g = 1$, $g_{1} =1$, $\gamma_{c} = 1$, $\delta_{c} = -2$, and $\delta_{g} = 0$ (dotted curve); 
$\delta_{g} = 1$ (dashed curve); $\delta_{g} = -1$ (full curve). }
\label{fig5}
\end{figure}

\begin{figure}
\centerline{\psfig{figure=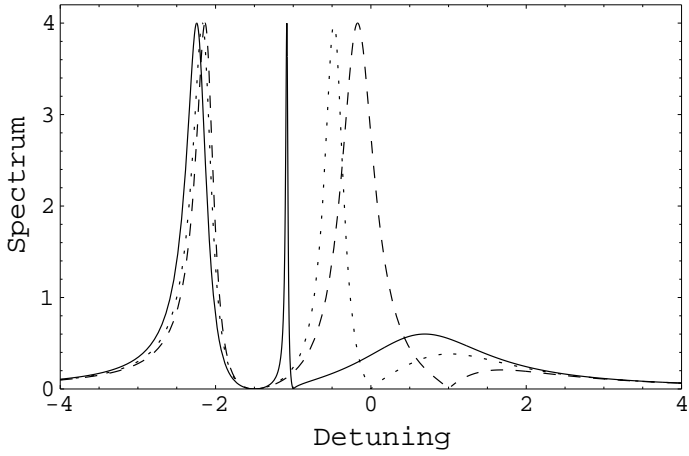,height=6.0cm}}
\caption
{\narrowtext The spontaneous emission spectrum $S(\delta_{\bf \lambda})$ 
given by Eq.\ (\ref{2spontiso})
for parameters 
$g = 1$, $\Omega = 1$, $\delta = -1.5$, $b_{2}(0)=1$, and $\delta_{g} = 0$ (dotted curve); 
$\delta_{g} = 1$ (dashed curve); $\delta_{g} = -1$ (full curve). 
All parameters are in units of $\gamma$. }
\label{fig6}
\end{figure}

In the case of an
isotropic photonic band gap material, with the use of 
Eq.\ (\ref{isoker}) we find,
\begin{equation}
S(\delta_{{\bf \lambda}}) \propto \gamma\left|\frac{(\delta_{\bf \lambda} - \delta)b_{2}(0) + \Omega b_{3}(0)}{(\delta_{\bf \lambda} - \delta)(\delta_{\bf \lambda} + i\gamma/2) + i g^{3/2}\frac{\delta_{\bf \lambda} - \delta}{\sqrt{\delta_{\bf \lambda} - \delta_{g}}} - \Omega^{2}}\right|^2 \, . \label{2spontiso}
\end{equation}
Let us assume first that the atom starts from its ground state, i.e.
$|b_{3}(0)|^{2}=1$, $|b_{2}(0)|^{2}=0$. 
Then, the spectrum obtains a dark line at $\delta_{\bf \lambda} = \delta_{g}$,
except in the case that $\delta = \delta_{g}$, so
in the case that the laser detuning becomes equal to the detuning 
from the band edge no dark line occurs, 
if the system is initially
in state $|3\rangle$. 
We note that in the case when the transition
$|2\rangle \rightarrow |1\rangle$ occurs in a Markovian
reservoir, then no dark line appears in the spectrum \cite{Zhu95a,Paspalakis98a}.
If now the atom starts in a superposition of the ground $(|3\rangle)$
and excited states $(|2\rangle)$ with real expansion coefficients $[b_{i}(0)]$
then two dark lines
can appear in the spectrum. The first dark line appears at $\delta_{\lambda} = \delta-\Omega b_{3}(0)/b_{2}(0)$ and is attributed to the laser-atom
interaction \cite{Paspalakis98a}. The second dark line occurs at
$\delta_{\lambda} = \delta_{g}$ and is attributed to the interaction
with the non-Markovian reservoir. In this case too, the second dark line disappears if
$\delta = \delta_{g}$.
The behaviour of the above spectrum for the case that the system
is initially in state $|2\rangle$ is shown
in Fig.\ \ref{fig6}. We have chosen $\delta_{g} \neq \delta$ therefore,
as predicted, two dark lines appears in the spectrum. 
We note the similarity of this spectrum and that of Fig.\ \ref{fig4}. 
This should be expected as the delta function density of modes 
(used in Fig.\ \ref{fig4}) represents a pure Jaynes-Cummings
interaction \cite{Knight93a}. 

\begin{figure}
\centerline{\psfig{figure=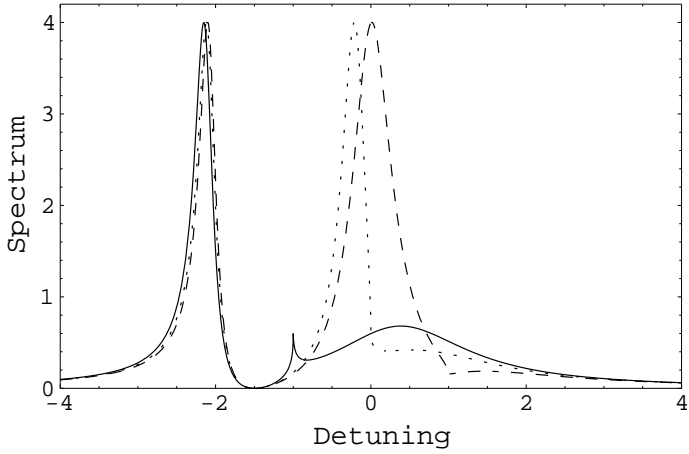,height=6.0cm}}
\caption
{\narrowtext The spontaneous emission spectrum $S(\delta_{\bf \lambda})$ given by Eq.\ (\ref{2spontsmo})
for parameters 
$g = 1$, $\epsilon = 0.3$, $\Omega = 1$, $\delta = -1.5$,  $b_{2}(0)=1$, and $\delta_{g} = 0$ (dotted curve); 
$\delta_{g} = 1$ (dashed curve); $\delta_{g} = -1$ (full curve). }
\label{fig7}
\end{figure}

If the more general, smoothed density of states of
Eq.\ (\ref{smodom}) is used (with $\epsilon \neq 0$), the
spontaneous emission spectrum is given by
\begin{equation}
S(\delta_{{\bf \lambda}}) \propto \gamma\left|\frac{(\delta_{\bf \lambda} - \delta)b_{2}(0) + \Omega b_{3}(0)}{(\delta_{\bf \lambda} - \delta)(\delta_{\bf \lambda} + i\gamma/2) + i g^{3/2}\frac{\delta_{\bf \lambda} - \delta}{i \sqrt{\epsilon} + \sqrt{\delta_{\bf \lambda} - \delta_{g}}} - \Omega^{2}}\right|^2 \, . \label{2spontsmo}
\end{equation}
Then, in the case that 
the atom starts in the ground state no dark line exists in the spectrum.
However, if the atom starts in a
superposition of the ground and excited states (with
real expansion coefficients) the laser-induced
dark line appears at the same frequency as before, i.e. at
$\delta_{\lambda} = \delta-\Omega b_{3}(0)/b_{2}(0)$. This is 
verified in Fig.\ \ref{fig7}, where only a single
dark line appears in the spectrum at $\delta_{\bf \lambda} = \delta$, as
$b_{3}(0) = 0$.

\begin{figure}
\centerline{\psfig{figure=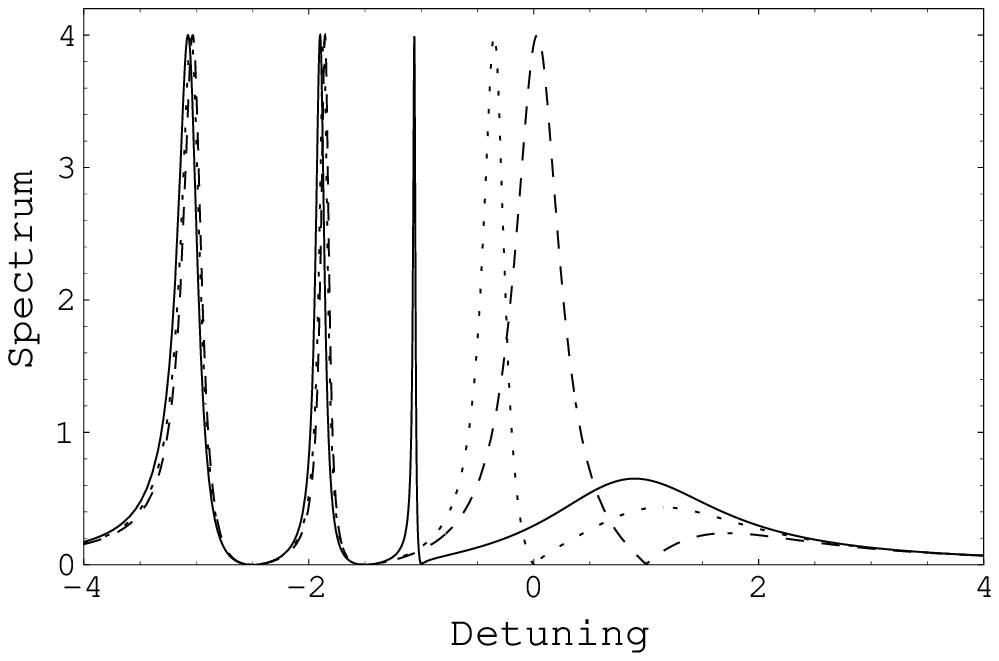,height=6.0cm}}
\caption
{\narrowtext The spontaneous emission spectrum $S(\delta_{\bf \lambda})$  given by Eq.\ (\ref{2spontdelta})
for parameters 
$g = 1$, $g_{1} =1$, $\delta_{c} = -2.5$,  $\Omega = 1$, $\delta = -1.5$,  $b_{2}(0)=1$, and $\delta_{g} = 0$ (dotted curve); 
$\delta_{g} = 1$ (dashed curve); $\delta_{g} = -1$ (full curve). }
\label{fig8}
\end{figure}

In the case of 
an isotropic photonic band gap with
a defect with delta function density of modes then the spectrum reads
\begin{equation}
S(\delta_{{\bf \lambda}}) \propto \gamma \left|\frac{(\delta_{\bf \lambda} - \delta)b_{2}(0) + \Omega b_{3}(0)}{(\delta_{\bf \lambda} - \delta)(\delta_{\bf \lambda} + i\gamma/2) + i g^{3/2}\frac{\delta_{\bf \lambda} - \delta}{\sqrt{\delta_{\bf \lambda} - \delta_{g}}}
- g^{2}_{1}\frac{\delta_{\bf \lambda} - \delta}{\delta_{\bf \lambda} - \delta_{c}}  - \Omega^{2}}\right|^{2} \, . \label{2spontdelta}
\end{equation}
and one obtains, in general, two dark lines 
if the atom starts from the ground state,
one at $\delta_{\lambda} = \delta_{g}$ and the other at $\delta_{\lambda} = \delta_{c}$.
Any of these dark lines can disappear if either $\delta = \delta_{g}$, 
or $\delta = \delta_{c}$ so the spectrum exhibits only
one dark line in this case. 
If now the atom starts in a superposition of the ground $(|3\rangle)$
and excited states $(|2\rangle)$ with real expansion coefficients
then a third dark line can appear in the spectrum at $\delta_{\lambda} = \delta-\Omega b_{3}(0)/b_{2}(0)$. The spectrum for the case that
the atom is initially in the excited state $|2\rangle$ (as in the previous cases), is displayed in Fig.\ \ref{fig8}. A very rich behaviour of 
the spectrum is find. The spectrum now has three dark lines at the predicted
values and four different peaks can be seen. 

\begin{figure}
\centerline{\psfig{figure=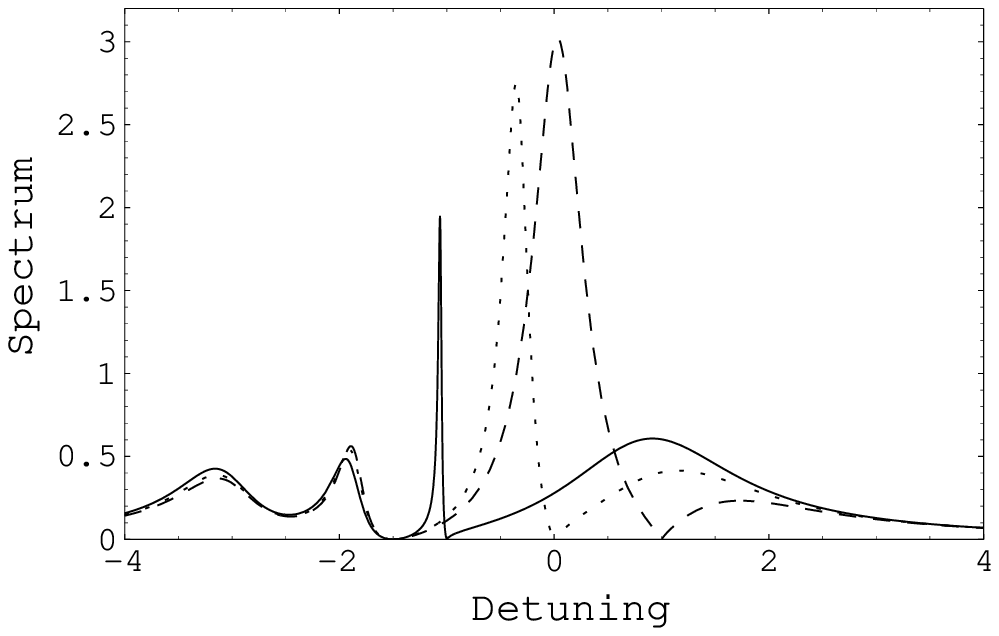,height=6.0cm}}
\caption
{\narrowtext The spontaneous emission spectrum $S(\delta_{\bf \lambda})$ 
given by Eq.\ (\ref{2spontlor})
for parameters 
$g = 1$, $g_{1} =1$, $\gamma_{c} = 1$, $\delta_{c} = -2.5$, $\Omega = 1$, $\delta = -1.5$,  $b_{2}(0)=1$, and $\delta_{g} = 0$ (dotted curve); 
$\delta_{g} = 1$ (dashed curve); $\delta_{g} = -1$ (full curve). }
\label{fig9}
\end{figure}

Finally, in the case of an isotropic photonic band gap with
a defect with Lorentzian density of modes the spectrum
gets the form,
\begin{equation}
S(\delta_{{\bf \lambda}}) \propto \gamma\left|\frac{(\delta_{\bf \lambda} - \delta)b_{2}(0) + \Omega b_{3}(0)}{(\delta_{\bf \lambda} - \delta)(\delta_{\bf \lambda} + i\gamma/2) + i g^{3/2}\frac{\delta_{\bf \lambda} - \delta}{\sqrt{\delta_{\bf \lambda} - \delta_{g}}} 
+ i g^{2}_{1}\frac{\delta_{\bf \lambda} - \delta}{i(\delta_{c} - \delta_{\bf \lambda})+\gamma_{c}/2} - \Omega^{2}}\right|^2 \, . \label{2spontlor}
\end{equation}
The dark lines in the spectrum
appear at the same frequencies as those of the simple  
isotropic photonic band gap model, discussed above. However, 
the shape of the spectrum in this case 
differs from that of Fig.\ \ref{fig6} as it is shown in Fig.\ \ref{fig9}.

\section{Summary}
In this article we have investigated the spontaneous emission properties
of two distinct atomic models with one transition
coupling to a Markovian reservoir while another transition
coupling to a non-Markovian reservoir. Of specific interest
to us were the existence of dark lines in the Markovian
spontaneous emission spectrum. We have shown that dark lines
can occur if the  non-Markovian reservoir
is described by certain densities of modes. In 
the case of the laser-driven
scheme of Fig.\ \ref{fig1}(b) laser-induced dark lines can co-exist
with non-Markovian reservoir-induced dark lines. Overall, 
spontaneous emission in the Markovian transition can be 
efficiently controlled  (and even suppressed) by appropriately
engineering the density of modes of the non-Markovian reservoir.

\section*{Acknowledgments}
E.P. would like to thank Niels J. Kylstra for useful discussions 
in the subject. 
The contribution of D.G.A. to this work was done in partial
fulfilment of the MSc.\ requirements at the Department
of Physics, University of Crete, Greece.
This work was funded by the UK Engineering and Physical Sciences Research Council (EPSRC) and the European Commission Cavity QED TMR Network ERBFMRXCT96066. D.G.A. wishes to
acknowledge the financial support of the 
Hellenic State Scholarship Foundation (SSF).


\pagebreak

\vspace{2.5cm}

\pagebreak

\vspace{2.5cm}

\vspace{2.5cm}

\pagebreak

\vspace{2.5cm}

\end{document}